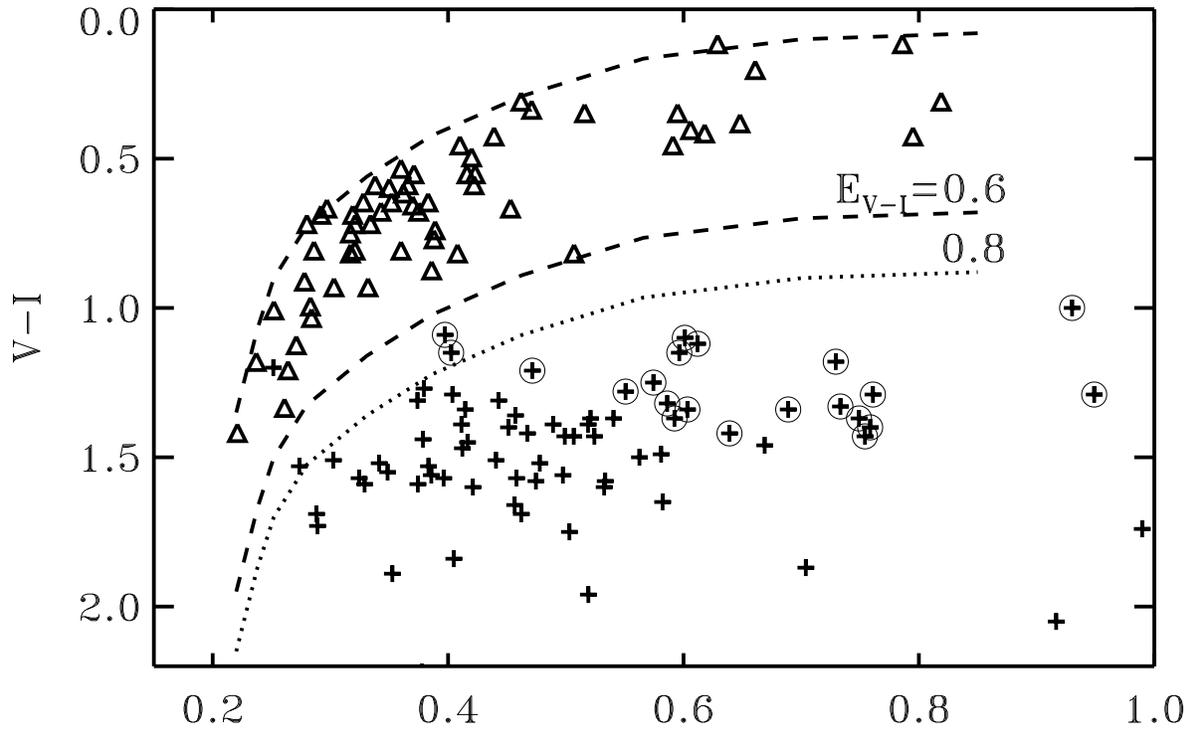
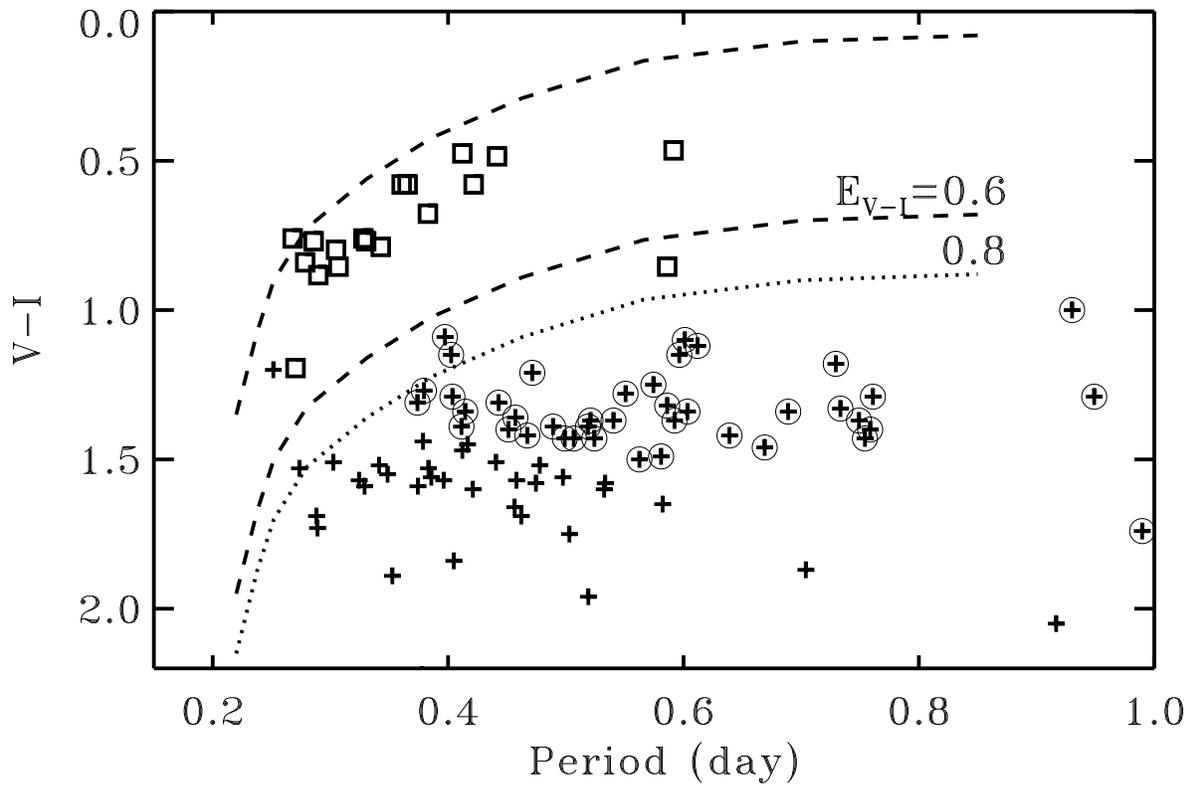

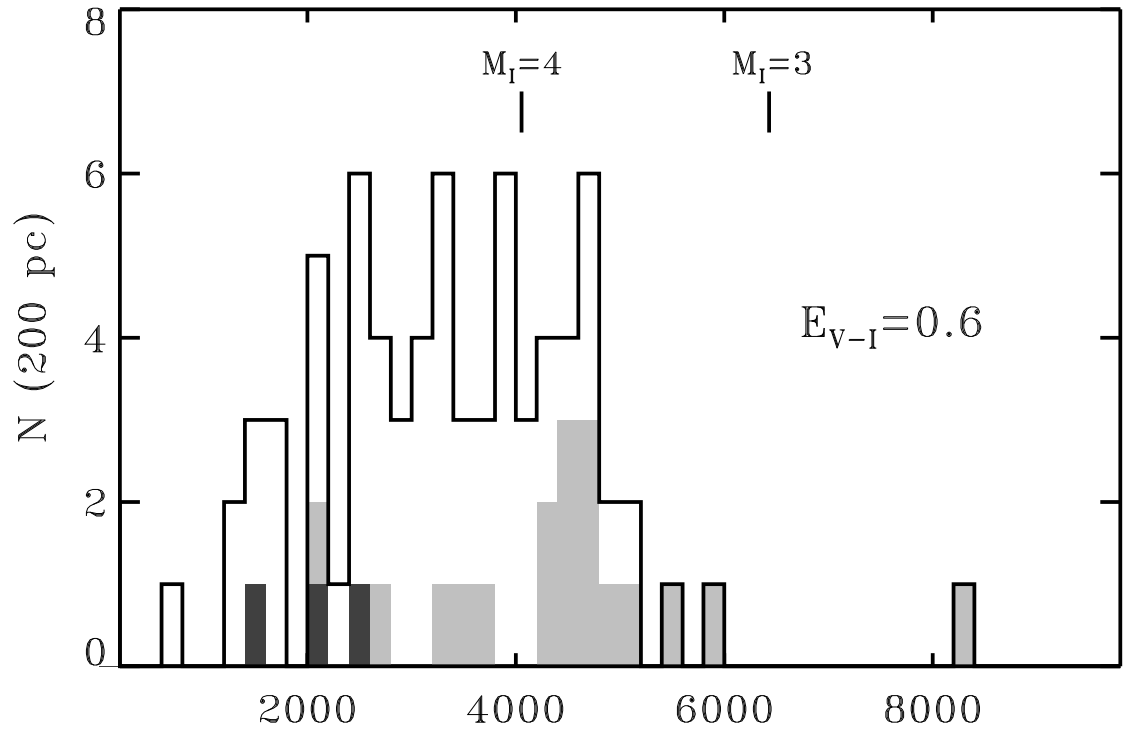
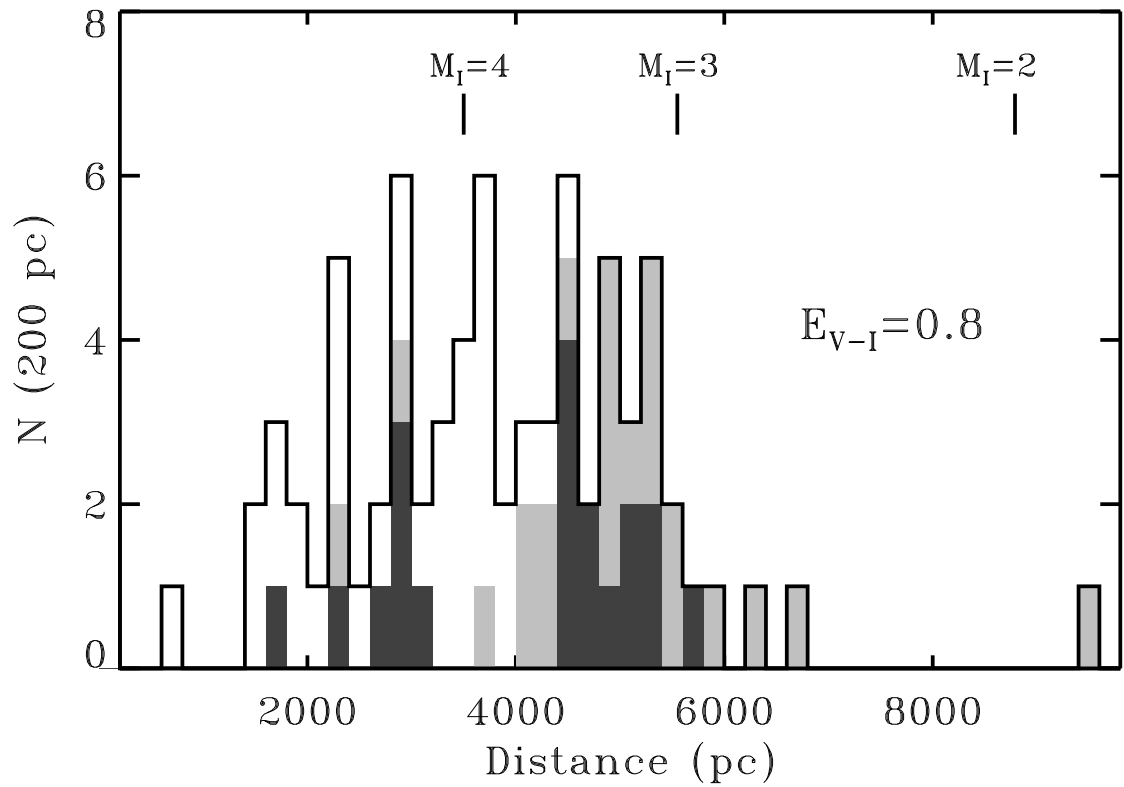

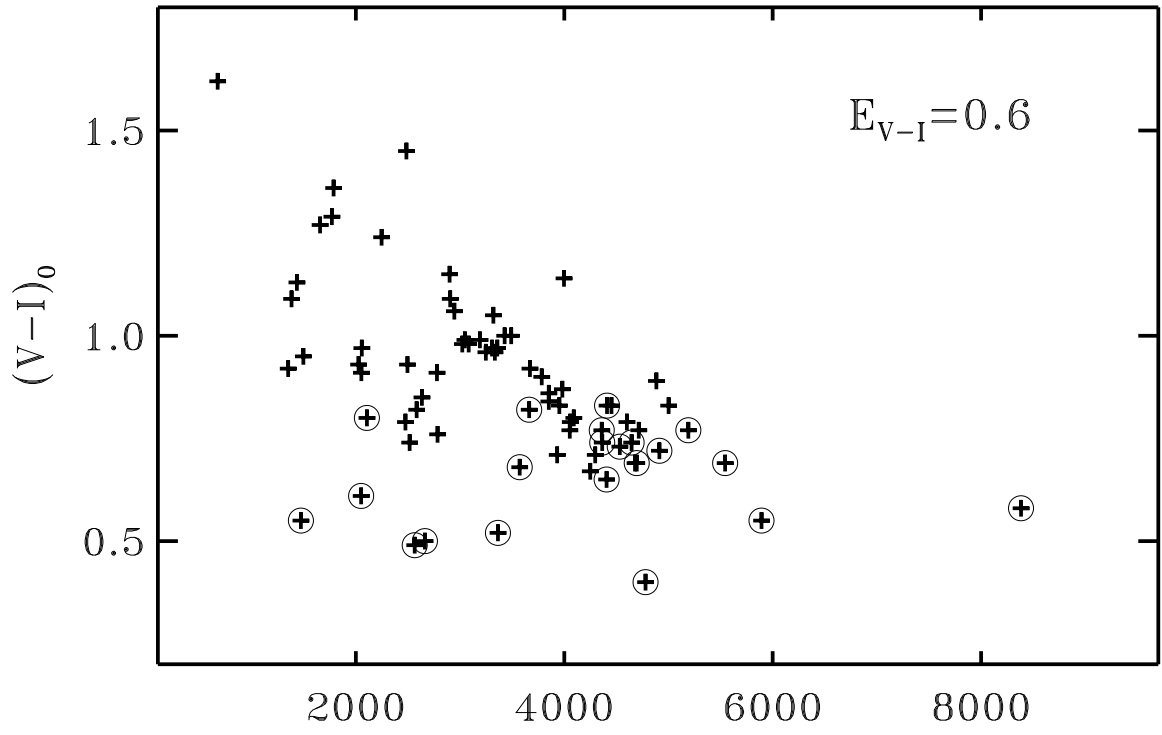
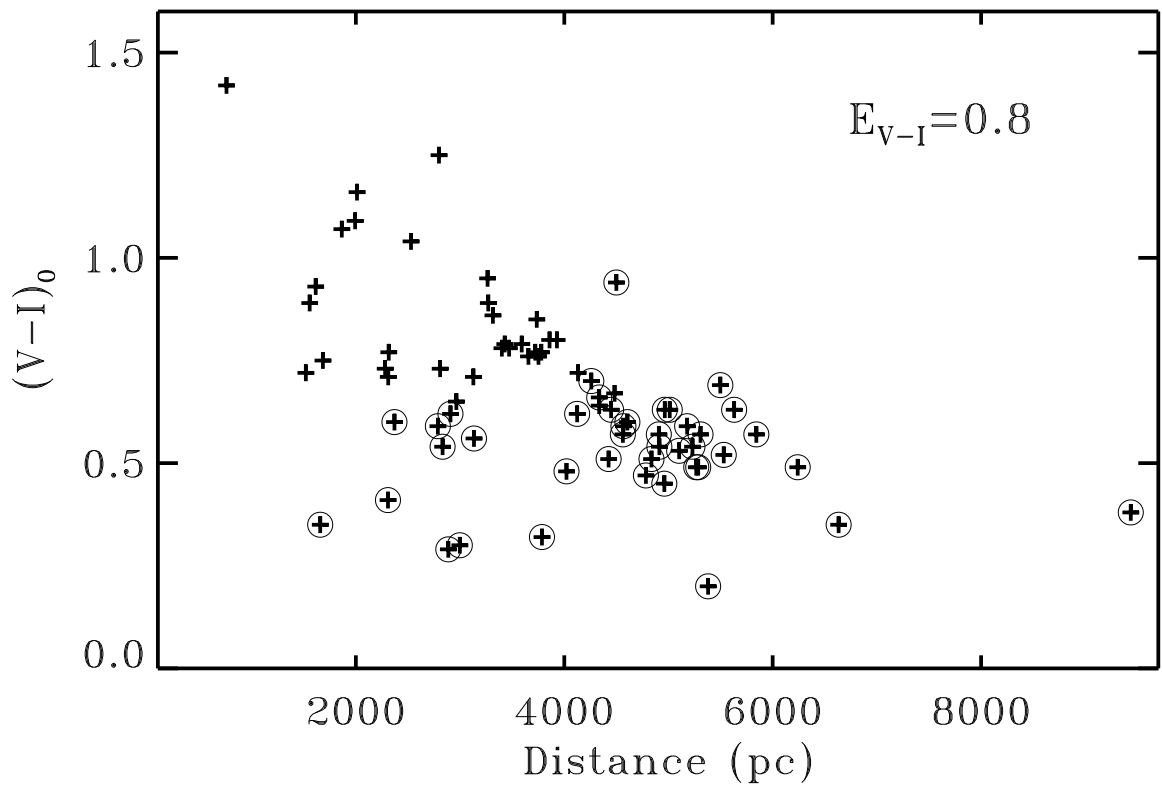

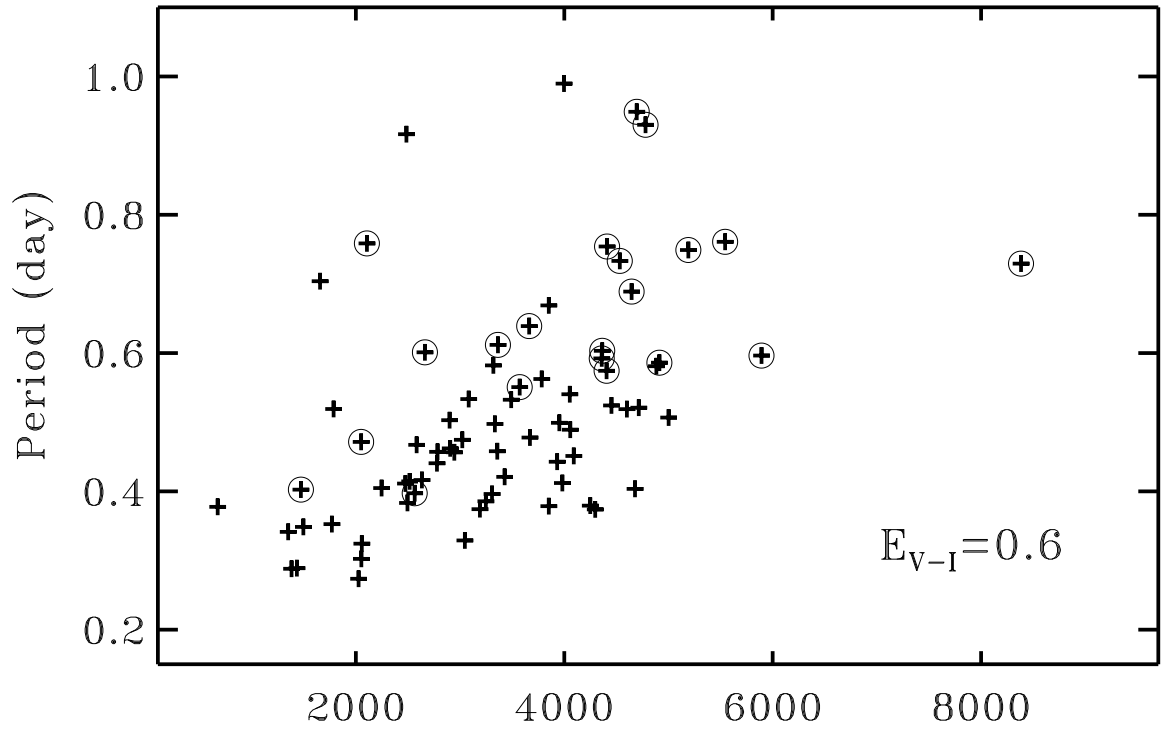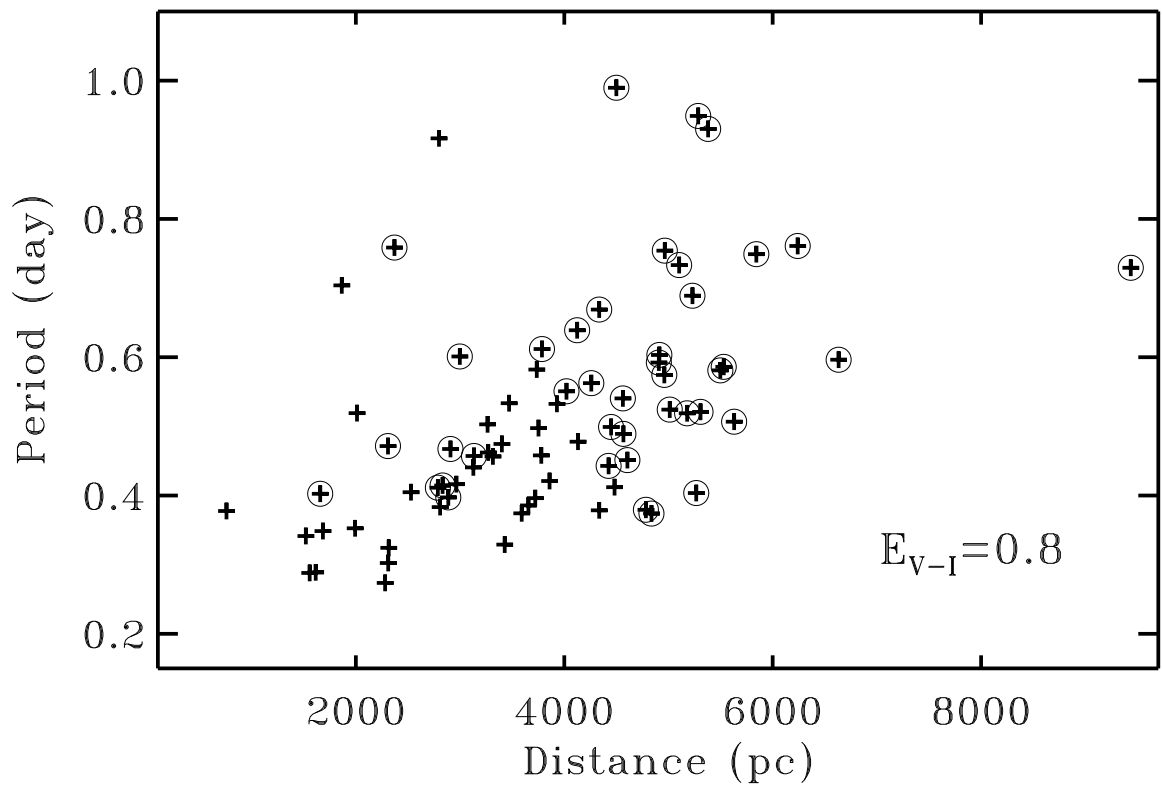

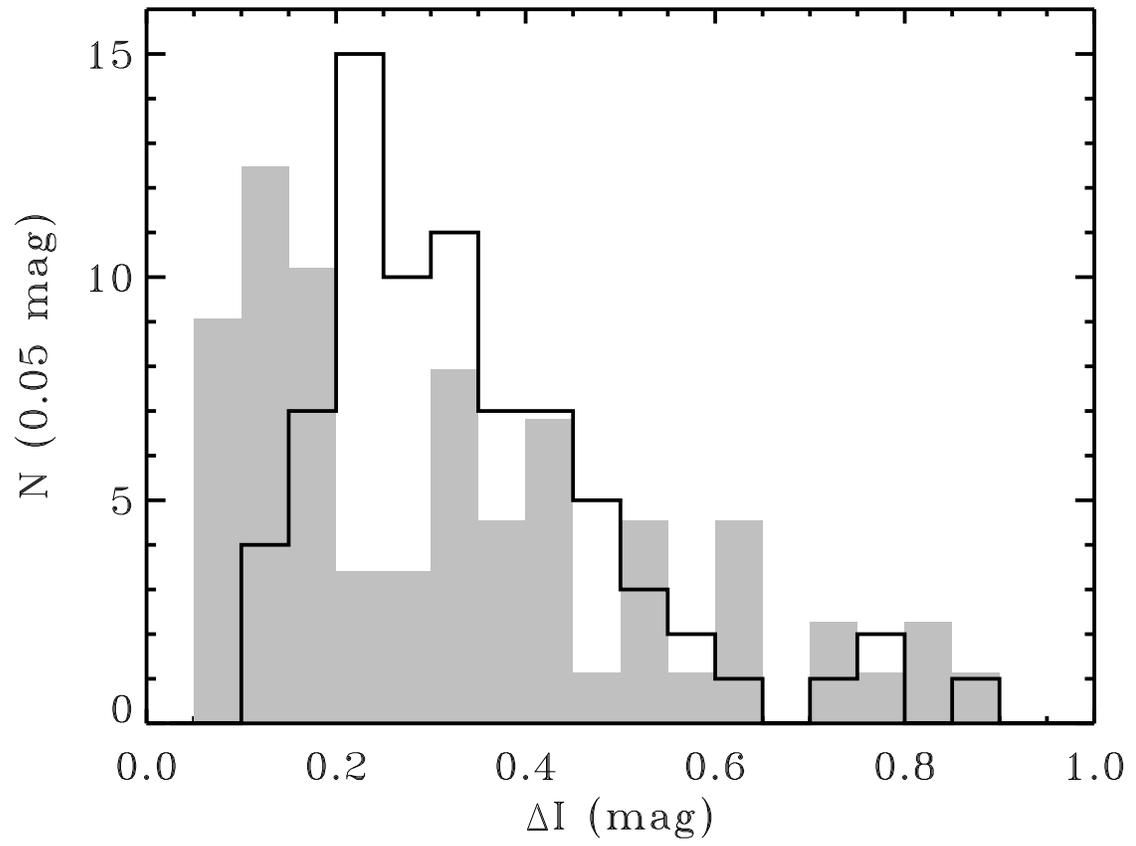

# W UMa-type Systems in the Central Baade Window Discovered in the OGLE Experiment

Slavek M. Rucinski[1]

e-mail: *rucinski@astro.utoronto.ca*

81 Longbow Drive, Scarborough, Ontario M1W 2W6, Canada

February 20, 1995

## ABSTRACT

77 W UMa-type systems to $I = 18$ mag in one of the 21 fields of OGLE have been assessed in terms of utility for extraction of astrophysical information from time-independent data such as colors, periods, maximum magnitudes, and variability amplitudes in the final global OGLE catalog which will contain $> 1000$ contact binaries. A subset of systems with absolute magnitudes $M_I < 3$ permits mapping of stellar distribution towards the Galactic Bulge to about 6 kpc from us, with indications that the reddening $E_{V-I} > 0.6$.

*Subject headings:* stars - stars: contact binary - stars: W UMa-type - galactic structure - gravitational micro-lenses

## 1. INTRODUCTION

The Optical Gravitational Lensing Experiment (OGLE) is a long term observing project aimed at detecting invisible matter in our Galaxy by detection of gravitational microlensing events in the direction of the Galactic Bulge. It has given a rich spectrum of by-product results, among them excellent color-magnitude diagrams (Udalski et al. 1993) which have provided interesting insight into the structure of the central parts of the Galaxy (Paczynski et al. 1994a (= P94), 1994b, Stanek et al. 1994).

This paper is the first utilization of another advantage of this dataset: its systematic temporal coverage permitting detection of very large numbers of periodic variable stars. The newly released first part of the variable star catalogue for the the central (BWC) Baade Window field (Udalski et al. 1994) contains 213 periodic variables to the limiting magnitude $I = 18$: 31 pulsating, 66 miscaleneous and 116 eclipsing; among the latter, 77 contact binaries of the W UMa type identified in the BWC Catalogue as EW-type. The goal of this paper is to analyze data for the contact binaries there, with an expectation of learning about advantages and limitations of the whole dataset in which BWC is one of 21 fields.

---

[1]Affiliated with the Department of Astronomy, University of Toronto and Department of Physics and Astronomy, York University

A simple extrapolation of the number of variable stars in BWC leads to a prediction that the final total catalogue will contain some 1000 – 1500 contact binaries. This estimate should be taken in perspective: The General Catalogue of Variable Stars lists 562 contact systems but for only 510 of them are the periods reliably known, whereas moderately good photometric data exist for some 130 systems, but only for half of that in standard photometric systems. Thus, the OGLE sample might tremendously enrich our knowledge of these systems; in the same time, relative ease of their detection and simplicity of their observational characteristics might substantially contribute to studies of the galactic structure.

Contact binary stars are still poorly understood in terms of their structure and origin (for a review, see Rucinski 1993). Recent systematic searches in stellar clusters (reviewed in Kaluzny & Rucinski 1993, Mateo 1993, Rucinski & Kaluzny 1994) have led to upward revisions of their frequency, both in clusters and in the galactic field. They are also recognized as extremely important for dynamical evolution of globular clusters (Hut et al. 1992).

The variable-star data for BWC consist of $V$ and (Cousins) $I$ magnitudes, amplitudes in $I$, periods, initial epochs and positions, as well as phased light curves in $I$. This paper is concerned with astrophysical information derivable from time-independent quantities. Analysis of light curves will lead to a separate publication.

## 2. PERIOD – COLOR RELATION

Relation between the period and color (PC) which is observed for W UMa-type systems is the most obvious tool to apply to the BWC systems. This relation is known to be only moderately tight as it reflects evolution of some contact systems leading to longer periods for more evolved systems. The short-period/blue envelope of the PC relation should be defined by the least-evolved systems, assuming that all systems have the same metallicity. For a metal-poor population, the left-edge is shifted to still shorter periods and bluer colors, as demonstrated by Rucinski (1994a) for contact systems with measured $B - V$ colors in globular clusters. For the $V - I$ color, this metallicity dependence is expected to be weaker than for the $B - V$ color. Neglecting relatively weak coupling between metallicity and period through the internal structure, we can estimate the direct influence on the color using the atmosphere models for solar-type stars by Buser & Kurucz (1992): $\Delta(B - V) \propto +0.08\,[Fe/H]$ and $\Delta(V - I) \propto +0.04\,[Fe/H]$. Whereas stars in the Galactic Bulge might show a spread as large as $-1 < [Fe/H] < +1$ (Frogel 1988), the Disk stars that we see in the current sample should have $-0.5 < [Fe/H] < +0.5$. Thus, we can expect $|\Delta(V - I)| < 0.02$.

At present, very few bright W UMa-type systems have measured $V - I$ colors. Among 18 systems used by Rucinski (1994b = R94) to establish the $M_V = M_V\,(log\,P,\,B - V)$ calibration, only two were observed in the $V - I$. A preliminary period-color relation for the $V - I$ color had to be therefore constructed assuming that it is legitimate to transform $B - V$ colors into $V - I$ colors using the Main Sequence relations. Such an assumption is debatable, as we do not know how high chromospheric activity of the W UMa systems might affect their color relations.

Figure 1 shows the PC relation for 75 systems in BWC with known $V - I$ colors, compared with two samples of bright W UMa systems. The first standard sample is that of Mochnacki (1985) who tabulated periods and dereddened B-V colors for 59 systems. The $V - I$ colors have been obtained using the Main Sequence relations of Bessell (1979, 1990). The second standard sample consists of 18 systems with independently known $M_V$ which were used in R94 for the absolute magnitude calibration. Some of these systems appear also in the sample of Mochnacki but their data have been independently derived from observations and assessed for consistency.

The interstellar extinction in the BWC region has been estimated to be about $A_V = 1.5 - 1.78$ in an application of the OGLE sample to study the galactic structure by Paczynski et al. (1994a = P94). With $A_V = 2.6\,E_{V-I}$ and thus $A_I = 1.6\,E_{V-I}$, the reddening is expected to fall within $0.58 < E_{V-I} < 0.68$. As we will see, most of the W UMa systems in BWC are located beyond 2 kpc, so that there was no need to consider the increase in $E_{V-I}$ to the assumed limiting value for distances smaller than 2 kpc (cf. P94).

The main results from application of the PC relation to the BWC systems are as follows (see Fig. 1): (1) Although some systems are close to the envelope at $E_{V-I} = 0.6$, most systems are located much below it; this might indicate that their reddening is larger that this value. We used 0.6 and 0.8 to estimate sensitivity of results to the assumption on the reddening. (2) The magnitude-limited character of the BWC sample results in an observational-selection effect of stretching the distribution of points horizontally in the PC relation: The distant, long-period, bright systems are over-represented in the BWC sample. (3) The reddening excess seems to be larger for systems with periods longer than about 0.5 day, but this cannot be taken as a sign of $E_{V-I}$ increasing with distance, as colors of intrinsically bright, long-period systems might be systematically modified by evolution. (4) One system at the short-end, V118 ($P = 0.2515$ d, $V - I = 1.20$), might be either a very nearby system with very small reddening or its period and/or variability type have been mis-classified. We note that its amplitude is only 0.14 and that its light curve indicates a more complex variability than normally seen in contact binaries.

## 3. ABSOLUTE MAGNITUDE CALIBRATION

An absolute $I$-magnitude calibration was needed to estimate distances to the W UMa systems in BWC. A preliminary version of such a calibration, based on the same 18 systems as in R94 with $B - V$ colors transformed to $V - I$ is given here. The least-squares solution of 18 equations for 3 unknowns in: $b_{P(VI)}\,log\,P + b_{VI}\,(V - I) + b_{0(VI)} = M_I$, gave: $b_{P(VI)} = -3.23 \pm 1.19$, $b_{VI} = +2.88 \pm 0.82$ and $b_{0(VI)} = -0.03 \pm 1.00$, with $\sigma = 0.29$. These equations were weighted uniformly since all the colors, except those for 44i Boo and VW Cep (which have uncertain $V - I$ measurements, see R94), were derived from the $B - V$ colors.

The same technique of bootstrap sampling, as in R94, was used to estimate uncertainties in $b$. The bootstrap sampling consists of repeated solutions (here 1000 times) of randomly sampled data with repetitions. It gave the following median and $\pm 1$-$\sigma$ ranges for the distinctly non-Gaussian distributions of the results: $b_{P(VI)} = -3.32^{+0.93}_{-2.39}$, $b_{VI} = +2.86^{+0.43}_{-0.96}$ and $b_{0(VI)} = -0.12^{+0.35}_{-0.34}$. Note that strong correlations between the coefficients result in smaller uncertainties in $M_I$ than these large errors in the coefficients would imply. In fact, extensive Monte Carlo experiments show that $M_I$ can be predicted to better than $\pm 0.2$ mag (typically $\pm 0.1$ mag) over the ranges of $0.27 < P < 0.59$ day and $0.46 < V - I < 1.19$ used in the calibration. This means that uncertainty in distance determinations is to a similar degree affected by uncertainties in the absolute magnitude as in the reddening (each at the level of about 15% in the distance).

Our main concern are systematic errors in the calibration, most notably that (1) it has been obtained through transformation of $B - V$ colors into $V - I$, and (2) that our calibrating stars had $P < 0.59$ day whereas a large fraction of systems seen at large distances have periods longer than this limit (we address this below by defining a "Good Bright Sample" among the BWC systems). The mediocre quality of the calibration, primarily due to the lack of standard $V - I$ data for W UMa systems of known distances and poorly understood properties of long-period contact systems are the main weaknesses of our approach.

## 4. W UMA SYSTEMS AS DISTANCE PROBES

Determination of distances was straightforward: $\log r\,(\mathrm{pc}) = (I - M_I - 1.6\,E_{V-I} + 5)/5$, with the absolute-magnitude calibration as given in Section 3, $M_I = M_I\,(\log P, (V - I) - E_{V-I})$. The assumed $E_{V-I}$ will be given in parentheses for values of $M_I$ or $r$. It should be noted that because the color coefficient in the $M_I$ calibration ($+2.88$) is larger in absolute value than the coefficient in the extinction law ($-1.6$), an increase in the assumed reddening leads to a non-intuitive increase in distances.

Histograms of derived distances are shown in Figure 2 for both assumptions on $E_{V-I}$. Most systems are located at distances between 2 and 5 kpc, but the upper limit results from the magnitude-limited character of the BWC sample. The distances reached for specific values of $M_I$ and for the assumed depth of the survey of $I = 18$, as estimated by Udalski et al. (1994), are marked in both panels of Figure 2. If we limit ourselves to only intrinsically bright systems, the sample obviously decreases in size. For $M_I < 2$, the sample consists of only one object for $E_{V-I} = 0.6$, and of 8 objects for $E_{V-I} = 0.8$. Among these systems the most luminous is V43 ($I = 15.58$, $V - I = 1.00$, $P = 0.9302$ d, $M_I(0.6) = 1.22$, $M_I(0.8) = 0.65$, $r\,(0.6) = 4790$ pc, $r\,(0.8) = 5380$ pc), whose brightness is due to a long period and moderately late spectral type. It is definitely an evolved object. Another bright system, V198, ($I = 17.66$, $V - I = 1.18$, $P = 0.7294$ d, $M_I(0.6) = 2.08$, $M_I(0.8) = 1.51$) is close to or within the Bulge ($r\,(0.6) = 8380$ pc,

$r\,(0.8) = 9440$ pc).

A larger sample with $M_I < 3$ (from now on called the "B-sample" for " Bright Sample"), consists of 21 systems for $E_{V-I} = 0.6$ and 41 systems for $E_{V-I} = 0.8$. Absolute magnitudes of these systems happen to be confined between $2.0 < M_I < 3.0$ and $1.5 < M_I < 3.0$, respectively, so that they should be good "standard candles". However, about half of the B-sample systems have periods longer than 0.59 day and thus their absolute-magnitude calibration is uncertain. To remedy this, we have defined a "Good B-sample" of bright systems with $P < 0.55$ day (the Monte Carlo experiments indicated that our calibration worsens at the long-period end). The histograms of distances for such systems are shown in Figure 2 in grey shade. They should provide good means for mapping of stellar distribution to distances of about 6 kpc. For both values of reddening the distance distributions truncate at or even before the 6 kpc mark, although for $E_{V-I} = 0.8$ the discrepancy is smaller. This argues for a larger value of $E_{V-I}$, as also suggested in P94, although the disagreement might be manifestation of a systematic error in the $M_I$ calibration for bright, long-period systems.

Perhaps the most interesting application of the distance distributions in Fig. 2 is to relate them to the striking increase in star numbers at about 2.5 kpc from us, with a sudden drop beyond that distance. Both features were suggested by Paczynski et al. (1994a) on the basis of the color-magnitude diagrams for stars in the OGLE fields. We seem to see a small peak at about 2 – 3 kpc, but this increase is defined by just a few systems. A substantial excess in numbers of bright W UMa systems is visible beyond about 4 kpc; however, it would be premature to state if this increase is discordant with the increase simply due to the $\propto r^2$ law. We are pleased to see that the main results for the Good B-sample do not differ from the B-sample; in fact, the distance clustering might be even better defined in it.

Barring interesting astrophysical phenomena (which cannot be excluded for changing stellar population at $|b| = 4°$ and distances $> 2$ kpc), the Bright Sample should not show any correlation between periods and distances, and between colors and distances. The correlation diagrams are shown in Figures 3 and 4, each for both values of reddening. The full sample clearly shows the expected correlations which are entirely explainable by a Malmquist-bias selection effect: at small distances we see mostly faint, red, short-period systems. However, the B-sample seems to contain systems appearing with roughly the same probability along the line of sight, irrespectively of period or color. The color correlation is definitely not present for the 19 and 41 systems of the Bright Sample (the Spearman rank correlation coefficients are $r_s = 0.08$ and $-0.02$ for the correlation "distance – de-reddened color", for both values of $E_{V-I}$). A weak correlation "period – distance" does seem to exist for the B-sample ($r_s = +0.52$ with significance level $> 98\%$ for $E_{V-I} = 0.6$, and $r_s = +0.40$ with significance level $> 99\%$ for $E_{V-I} = 0.8$) but this correlation is entirely due to systems which do not belong to the Good B-sample.

We stress that the usefulness of the OGLE sample will increase very substantially with the availability of data for all 21 fields as then the short-period systems, which are under-utilized here because of small numbers, could be used as excellent probes to distances < 4 kpc. Also, we note that for an increased limiting magnitude to $I = 19$, the full analogue of the B-sample will reach as far as into the Bulge.

## 5.  VARIABILITY AMPLITUDES

Distribution of amplitudes $\Delta I$ of the W UMa-type systems in BWC has been compared with that for systems discovered recently through CCD searches in open clusters (Kaluzny & Rucinski 1993, Rucinski & Kaluzny 1994), the latter sample avoiding the strong bias against small amplitude systems which is clearly visible in the sky-wide sample of bright W UMa systems. These samples are roughly similar in size, 68 in the clusters versus 77 here. We compare distributions of observed amplitudes in Figure 5. The figure indicates that the distributions differ significantly for amplitudes smaller than 0.3 mag. It is possible that the threshold for variability and/or periodicity detection was higher than was claimed by Udalski et al. (1994), i.e. probably 0.2 mag, and not at 0.1 mag. If this was the case, re-normalization of distributions could probably produce agreement. We should note that the BWC sample is probably more complete in the period coverage than the open-cluster sample, the latter being based on observations usually obtained during short observing runs and thus probably under-representing long-period systems. For long periods, we can expect a stronger admixture of evolved systems with dissimilar components which show small amplitudes. This might explain the disagreement of distributions in the 0.2 – 0.35 magnitude region.

We note that the distribution of variability amplitudes, $A(a)$ (for OGLE, $a = \Delta I$), has a potential of giving a distribution of mass-ratios, $Q(q)$, through solution of the integral equation: $A(a) = \int K(a,q)Q(q)dq$, where the kernel $K$ can be calculated for randomly oriented orbits. Such a solution utilizes the known decrease of amplitudes for mass-ratios different from unity. Experiments with artificial data and with distributions for systems in open clusters show that about 500 to 1000 systems are needed for meaningful results on $Q(q)$ so that the present sample is too small, but the full OGLE Catalogue should give sufficient data.

Most sincere thanks are due to Bohdan Paczynski for his unlimited readiness to comment, suggest and criticise, and to my wife for support of my research. I also thank the OGLE team for making all data available over the computer network.

The research grant from the Natural Sciences and Engineering Council of Canada used to pay the page charges is acknowledged with gratitude.

– 7 –

---

Figure captions

1. The period–color relation for 75 W UMa-type systems in BWC with measured $V-I$ colors (crosses). The upper panel compares the BWC data with 59 bright, sky-field systems whose $B-V$ colors (Mochnacki 1985) have been transformed to $V-I$ (triangles). These systems have been used to draw an approximate short-period/blue envelope of unevolved contact binaries (broken line). The envelope is shown shifted by $E_{V-I} = 0.6$ and 0.8 in both panels. The lower panel compares the BWC systems with a more restrictive sample of 18 systems used for the absolute-magnitude calibration (R94). In both panels the data points of the B-sample (see below) of systems with absolute magnitudes $M_I < 3$ have been marked by circles around the crosses.

2. Histograms of distances for the whole sample (line) and for the B-sample of systems with $M_I < 3$ (grey shade) for two assumed values of the reddening. The Good B-sample ($P < 0.55$ day) is distiguished by a darker shading. Distances corresponding to the limiting magnitude of the BWC sample of $I = 18$ and various absolute magnitudes are marked in the upper parts of both panels.

3. The correlation diagram between distances and de-reddened colors for the BWC systems (crosses). The B-sample systems are marked by additional circles around the crosses.

4. The correlation diagram between distances and periods of the BWC systems. The symbols are the same as in Figure 3.

5. Variability amplitude distributions for the BWC sample (line) and for W UMa systems in open clusters (grey shade; Kaluzny & Rucinski 1993, Rucinski & Kaluzny 1994). The latter has been re-normalized to the same number of systems as in the BWC sample.